# Radiation resistance in simulated metallic core-shell nanoparticles

D.R. Tramontina[1], O.R. Deluigi[1], R. Pinzón[2-5], J. Rojas-Nunez[6,7], F.J. Valencia[7,8], R.C. Pasianot[9], S.E. Baltazar[6,7], R.I. Gonzalez[7,10], E.M. Bringa[1,10*]

[1] CONICET and Facultad de Ingeniería - Universidad de Mendoza, Mendoza, 5500, Argentina.
[2] Universidad Tecnológica de Panamá, Centro de Investigaciones Hidráulicas e Hidrotécnicas (CIHH), República de Panamá.
[3] Sistema Nacional de Investigación (SNI-SENACYT), República de Panamá.
[4] Grupo CIHH del HPC-Cluster-Iberogun, República de Panamá.
[5] Centro de Estudios Multidisciplinarios de Ingeniería Ciencias y Tecnología (CEMCIT-AIP), El Dorado, República de Panamá.
[6] Departamento de Física, Universidad de Santiago de Chile, Santiago, Chile.
[7] Center for the Development of Nanoscience and Nanotechnology (CEDENNA), Chile.
[8] Departamento de Computación e Industrias, Facultad de Ciencias de la Ingeniería, Universidad Católica del Maule, Talca, Chile.
[9] Gerencia Materiales CNEA, CONICET and Instituto Sabato UNSAM/CNEA, Buenos Aires, Argentina.
[10] Centro de Nanotecnología Aplicada, Facultad de Ciencias, Universidad Mayor de Chile, Santiago, Chile 8580745.
Corresponding author e-mail: ebringa@yahoo.com

## Abstract

We present molecular dynamics (MD) simulations of radiation damage in pure Fe nanoparticles (NP) and bimetallic FeCu core-shell nanoparticles (CSNP). The CSNP includes a perfect body centered cubic (bcc) Fe core coated with a face-centered cubic (fcc) Cu shell. Irradiation with Fe Primary Knock-on Atoms (PKA) with energies between 1 and 7 keV leads to point defects, without clustering beyond divacancies and very few slightly larger vacancy clusters, and without interstitial clusters, unlike what happens in bulk at the same PKA energies. The Fe-Cu interface and shell can act as a defect sink, absorbing radiation-induced damage and, therefore, the final number of defects in the Fe core is significantly lower than in the Fe NP. In addition, the Cu shell substantially diminishes the number of sputtered Fe atoms, acting as a barrier for recoil ejection. Structurally, the Cu shell responds to the stress generated by the collision cascade by creating and destroying stacking faults across the shell width, which could also accommodate further irradiation defects. We compare our MD results to Monte Carlo Binary Collision Approximation (BCA) simulations using *SRIM* code, for the irradiation of an amorphous 3-layer thin film with thickness equal to the CSNP diameter. BCA does not include defect recombination, so the number of Frenkel pairs is significantly higher than in MD, as expected. Sputtering yield (Y) is underestimated by BCA, which is also expected since the simulation is for a thin film at normal incidence. We also compare MD defect production to bulk predictions of the analytic *Athermal Recombination Corrected Displacements Per Atom* (arc-dpa) model. The number of vacancies in the Fe core is only slightly lower than arc-dpa predictions, but the number of interstitials is reduced by about one order of magnitude compared to vacancies, at 5 keV. According to the radiation resistance found for FeCu CSNP in our simulations, this class of nanomaterial could be suitable for developing new radiation resistant coatings, nanostructured components, and shields for use in extreme environments, for instance, in nuclear energy and astrophysical applications.

# 1. Introduction

There is a need to find radiation resistant materials to meet the demands of future nuclear energy technologies [1-3], with sustained efforts expected for nuclear fusion [4-6]. Enhanced resistance is also required for radiation shielding in space applications [7,8]. Significant work has been done in simulations and experiments using nanostructured materials to improve radiation resistance [2,9-13]. The simulation of irradiation-induced defects in single element metals like Cu and Fe using molecular dynamics (MD) is a mature field [14-17]. The general behavior involves a ballistic stage of the collision cascade, with defect production reasonably described by binary collision approximation (BCA) simulations [16,17]. Defects are usually quantified by the number of vacancy-interstitial pairs (Frenkel pairs, FP), reaching a maximum number $FP_{max}$ usually before 1 ps after irradiation. This initial stage is followed by defect recombination, and the final number of defects is significantly lower than the peak value, with around 70% of the defects recombining for 5 keV cascades in bulk Fe [18]. In Fe, for energies larger than ~5 keV, there is a thermal spike regime where some material is molten under non-equilibrium conditions. Self-interstitial atoms (SIA) are produced at the boundary of the cascade, and they can form SIA clusters. Usually, vacancies are created at the cascade center, and for Fe, they remain as single vacancies or form small vacancy clusters, in contrast with cascades in fcc Cu, where vacancy clusters can be large and lead to stacking fault tetrahedra (SFT) [15,16]. Most studies focus on high energy cascades, where SIA clusters lead to the formation of dislocation loops [19,20].

Lee *et al*. [21] simulated 2 keV cascades in FeCu, with 5% Cu, using a 2NN MEAM potential. This potential considered the mixing enthalpy, the composition dependencies of the lattice parameters in solid solutions, and the formation energies of point defects, but close range corrections, like the usual Ziegler-Biersack-Littmark (ZBL) potential, were not included. Zhang *et al*. [22] employed the EAM potential for FeCuNi from Bonny *et al*. [54] and added ZBL corrections below 0.15 nm to simulate collision cascades in Fe-Cu alloys with Primary Knock-on Atoms (PKA) energies of 2-20 keV. In this work the concentration of solute and its type did not impact the number of survival Frenkel pairs, and Cu produced a very small number of interstitials.

As for irradiation including surfaces, Granberg *et al*. [23] studied the formation of vacancy clusters and SIA loops in bulk Fe and Fe with a surface, which acted as defect sinks. Sand *et al*. [24] found that vacancy and interstitial cluster sizes follow scaling laws in both bulk and thin foil irradiation. Also, they observed the formation of craters when the liquid core of the heat spike extends to the surface, ejecting large amounts of material in the form of sputtered atoms and atom clusters.

There is an increasing number of simulations dealing with radiation damage in nanostructured materials [10,25-27]. The unifying principle is that interfaces and surfaces will act as defect sinks and reduce radiation damage. There are "bulk" nanostructured materials, like nanocrystals [11,28], nanolayered materials [10,29,30], and materials with porosity [9,12,31], where interfaces and internal surfaces can capture defects. There are many efforts to irradiate finite-size systems, like nanofilms [19,32], nanowires [27,33], and nanoparticles [26], since surfaces will act as defect sinks. Regarding Fe thin films, it was found that rapid interstitial migration to the film surface impeded the formation of the large interstitial loops observed in bulk irradiation [19].

Concerning irradiation effects on nanoparticles (NP), there are several studies on sputtering [34-36], generally finding enhanced sputtering when the NP diameter is comparable to the ion range. Sputtering can be greatly influenced by thermal spike events [36]. Greaves *et al*. carried out experiments and simulations on 80 keV Xe irradiation of Au nanorods, finding huge sputtering yields from thermal spike events [37]. Although most of the studies are on metallic NP [38], there are studies on irradiation effects in semiconductor NP, including vacancy formation [39]. Oxide NP were significantly more radiation tolerant than micron-sized particles [40].

Aradi *et al.* [41] found radiation tolerance under continuous He irradiation at 15 keV in W nanoparticles, finding no evidence for nucleation of He bubbles nor dislocations for NP nanoparticle radius below 20 nm. Aradi *et al.* [42] also studied the irradiation of steel NP with 300 keV ions experimentally and found the formation of a Cr-rich outer oxide layer. The resulting core-shell structures were radiation resistant for NP radii less than 25 nm, and NP with radius less than 10 nm displayed NP coalescence. Khan *et al.* [43] have studied Fe NP bombardment by Fe, as a function of bombarding energy, in the range 5-40 keV, with MD simulations. They used NPs of 15 nm diameter and found very few surviving FP.

In this study, we are interested in bimetallic core-shell nanoparticles, where experimental synthesis processes can be accurately performed and allow for the control of a variety of structures [44,45]. This is the case for FeCu CSNP [46]. Core-shell NP have two types of sinks; the NP outer surface and the interface between core and shell. These sinks would satisfy roles somewhat similar to the filament surfaces in nanofoams [9,12], the interfaces between layers for nanolayer composites [10,30], or between NP and matrix, as in complex alloys [47] or in Oxide Dispersion-Strengthened (ODS) Steels [13]. The effect of a silica shell has been studied for Au NP irradiation with a large keV He ions flux, using experiments and Monte Carlo simulations [48]. It has been shown experimentally that the core-shell interface in CSNP might be an effective sink for radiation-induced defects [49], and a coating made with a collection of such NPs might be a suitable material for high radiation environments. Recently, it was also shown that a bulk metallic glass (BMG) shell significantly improved the radiation tolerance of Au cubic NP [50]. This also opens the possibility of employing irradiation as a customizable nanostructuring tool for manipulating mechanical and chemical properties of core-shell nanoparticles, possibly enabling additional technological applications, like improved hardness [51] or enhanced chemical reactivity [52].

Most of the current simulations on bombardment of metallic NP have focused on sputtering. In this work, MD simulations have been used to study the irradiation of Fe and FeCu NP, and these results are compared to analytic models and related Monte Carlo simulations, as described in detail below.

## 2. Methods

Molecular dynamics (MD) simulations have been carried out with the LAMMPS code [53] using the microcanonical ensemble for an initial temperature of 10 K. Atomic interactions are defined by a multicomponent embedded atom method (EAM) interatomic potential for the ternary FeCuNi system developed by Bonny and coworkers [54]. All interactions (but Fe-Fe that already included the feature [55]) were smoothly matched to the universal ZBL function [56] at short distances, below about 0.2 nm, in order to allow for the simulation of collision cascades, at the same time avoiding changes in equilibrium and defects properties. The merging was effected by means of cubic splines, in a similar way as in our previous work [57], aiming at obtaining threshold displacement energies as close as possible to literature values [58]. Potential tables will be available from the NIST database [NIST:http://www.ctcms.nist.gov/potentials/index.html]. Simulations have also been carried out using the alternative ZBL spline by Zhang et al. [22].

Visualization and post-processing were done with OVITO [59]. Wigner-Seitz (WS) cells are employed to characterize interstitials and vacancies [60]. OVITO allows for affine transformations of the original reference lattice, something similar to the proposal by Mason *et al.* [61], improving the accuracy of Wigner-Seitz detection. For the classification of the local atomic structures we adopted the Common Neighbor Analysis (CNA) algorithm [62] and the Polyhedral Template Matching (PTM) method [63].

The target is a spherical Fe body-centered cubic (bcc) single crystal NP, with a ~7 nm diameter and ~15500 atoms. The z-axis is oriented along the [001] direction. The bimetallic CSNP, uses the Fe NP as core adding a face-centered cubic (fcc) single crystal Cu shell 1 nm thick, giving a CSNP with ~9 nm diameter, and a total of ~31000 atoms. The CSNP was constructed following the temperature ramp proposed by Rojas et al. [46]. The resulting CSNP was then subjected to a relaxation at 300 K, for 50 ps with a timestep of 1.0 fs. The CSNP contains some stacking faults in the Cu shell, prior to irradiation, as seen in Fig. S1. Localized damage in the CSNP was calculated considering several fixed regions of interest: the core with a radius of 3.2 nm; the shell starting from r >= 3.8 nm; and the interface region in between. This allows for an approximation neglecting topological irregularities.

A collision cascade is initiated by a primary knock-on atom (PKA), with an energy $E_{PKA}$= 1, 3, 5 and 7 keV. adopted. An adaptive timestep between 0.01 to 1.0 fs was adopted during the simulation to ensure energy conservation. The irradiation with the resulting defect production and recombination was studied until simulations times of 10 ps in the microcanonical ensemble. The Fe PKA is chosen near the top surface of the core, closer than in the recent setup used for a flat surface where PKA depths of ~6-12 nm were used [23]. The PKA velocity points downwards, in a randomly defined direction inside a cone, with an aperture angle of 45 degrees. To obtain statistically meaningful data, tens of cascades are required [23,64]. For this work at least 50 MD runs with random positions and velocity directions for the PKA were performed for each system, and reported MD values include the mean and standard error from these cases. There are many studies of NP bombardment at high energies [26,36,37], where the thermal spike phase of the cascade is important for defect accumulation and sputtering. A maximum in the sputtering yield could be expected below 10 keV, but energies below that have rarely been explored [26,65]. One PKA for the simulated nanoparticle would give an equivalent ion fluence close to $10^{14}$cm$^{-2}$. The NPs are surrounded by vacuum and placed in the middle of a cubic MD box. When sputtered atoms reach the side of the MD simulation box they are frozen in place, to avoid increasing box size and computational cost. The electronic stopping contribution is small for the energy range studied [56,66] and was neglected in the simulations. For reference, bulk irradiation simulations were also carried out.

Monte Carlo (MC) simulations [66] were also performed for a single comparison with our MD results, by use of the Binary Collision Approximation (BCA) method. A detailed description of those simulations is available in the Supplementary Materials section.

Many FP generated during the early stages of the collision cascade, as in BCA simulations, will actually recombine and disappear. A defect recombination rate $(\eta)$ can be defined as $\eta$=1-(FP$_{prim}$/FP$_{max}$), where FP$_{prim}$ is the number of FP at the end of primary damage, i.e. at the end of MD simulations, and FP$_{max}$ is the peak number of FP.

The arc-dpa [58] is also used to compare with our MD results. This model corrects the Norgett-Robinson-Torrens (NRT) damage model [71] with a prefactor, but requires two additional parameters, b and c:

$$N_{ARC-DPA} = \epsilon(T_d)\, FP_{NRT}, \text{ where } \epsilon(T_d) = \left[\frac{(1-c)}{\left(\frac{2E_d}{0.8}\right)^b}\right] T_d^b + c \qquad \text{(Eq. 1)}$$

$N_{ARC-DPA}$ is the surviving number of FP after primary damage, and $T_d$ is the deposited energy. The parameters we use for Fe are b=−0.568, c=0.286 [58]. Given that there is no electronic loss, we assume that $T_d$= $E_{PKA}$, but this is not entirely true for our NP simulations, where the range of some PKA exceeds the NP diameter. We can only apply this model to the pure Fe NP.

In addition to single PKA simulations, cumulative irradiation was performed for a PKA energy of 3 keV, with a total of 50 irradiations for both Fe the NP and CSNP. Each PKA atom was randomly selected in the core region of the NP, and given a velocity with directions chosen from the same set of 50 random directions from the single PKA simulations. Each cascade is simulated with an adaptive timestep during ~12 ps, followed by a relaxation during 5 ps, to allow for some defect relaxation between cascades, followed by the next PKA. To avoid NP rotation and displacement due to the successive bombardment, a group of 100 atoms are tethered to an anharmonic potential with the form $U=k|r-r_0|^3$, with $k=10$ eV/A³, and $r_0$ the initial position of each atom in the group. The atoms experiencing this restitutive force are located close to the NP top surface, and do not interact with the PKA atom nor the atoms in the subsequent collision cascade. In addition to this, before the Wigner-Seitz procedure to identify defects, we applied an affine rotational transformation to the final configuration, using non-damaged atoms as a reference, in order to minimize the effects of NP rotation due to momentum transfer after many PKA. This is done in the same spirit as the method proposed by Granberg *et al*. [23], which applies such transformations locally.

## 3. Results and Discussion

Cascades in the NP and CSNP lead to defect formation. Comparison of MD results and Monte Carlo simulations is included in the Supplementary Material. Fig. 1 shows defects of CSNP, and Fig. S2 also shows some typical cascade cases. The cascade leads to several replacement collision sequences, but they generally end-up at the surface of the Fe core or NP, and are not displayed as trajectories with the filtering used for Fig. 1.

Fig. 2 shows details for a 5 keV cascade, including dislocations for the CSNP Cu shell. We note that cascades in bulk Fe with the range of energies studied in this work produce mostly point defects, but also some self-interstitial atoms (SIA) clusters, as a result of primary damage [14,23]. However, loops are never observed in our simulations, partly due to surface defect-sinks and to ejected atoms which lower the energy deposited in the target. Higher energy might lead to defect clusters [19], mainly SIA clusters because vacancies in Fe are typically isolated or forming very small clusters [15]. Note that Fe interstitials are dumbbells for both the NP and the core of the CSNP.

We observe only a few Fe atoms in the Cu shell as result of the collision cascade, as shown in the examples in Fig. 1 and 2. Higher energy or higher dose might provide higher mixing induced by radiation [16]. A few energetic Fe atoms do cross the Cu shell and remain on the NP surface, also visible in the surface of the sample shown later in Fig. 8 at the left. No intermetallic phase is expected for pure FeCu [72], but the presence of impurities might lead to interfacial segregation. Cu atoms in a FeCu alloy were shown to have only weak interactions with radiation induced vacancy clusters [73]. Here, for the thin simulated Cu shell, the Fe-Cu interface does allow for damage mitigation in the pure Fe core. Since the Cu shell is thin, there is no significant clustering of vacancies, as in bulk irradiated Cu simulations [15].

Experimental CS nanocube irradiation with high energy ions leads to increased surface rugosity [50], while our single low energy cascades do not generally lead to surface roughness nor surface area increase, except for a few cases where the PKA exits the NP and leaves a small shallow crater behind.

Collision cascades can lead to pressure waves which, despite being extremely short, have been identified with shock waves [74,75]. In this simulations, stress can be efficiently dissipated by the NP surface, but the transient local stress produced by the cascade leads to the creation and destruction of stacking faults (SF), adjacent to the pre-existing stacking faults in the Cu shell of the CSNP, but without significant change in the SF content of the CSNP, which is shown in Fig. S1 for the unirradiated CSNP. Fig. S3 shows that atomic stress in the CSNP does not reach a transient value larger than ~10 GPa for a 7 keV PKA. The bcc Fe core would require higher stress to deform plastically, for instance, by twins [76]. The stress at the FeCu interface can enhance defect sink strength and hinder the formation of migration paths for defect mobility [27]. Experiments by Chen *et al.* showed that coherent immiscible interfaces in FeCu, in the limit of thickness smaller than 5 nm, are effective in mitigating radiation induced damage [29].

Fig. 3 shows the averaged evolution of defects for the simulations at 5 keV for the PKA. Vacancies are plotted in Fig. 3(a,b) and interstitials Fig. 3(c,d). We separate the contribution of different regions of the nanoparticles as core, surface or interface, shell and total volume (NP, CSNP). We observe the heat spike phase, then the recombination phase, and finally the beginning of the steady state phase in both samples. Same analysis for 3 keV PKA is shown in Fig S4 in the Supplementary Materials Section. The amount of defects in Fe is considerably reduced for the CSNP. In the early stages of cascade evolution, there are similar numbers of vacancies and interstitials, as expected. They start differing even before peak damage is reached, because many interstitials migrate and are absorbed by the FeCu interface. Study of atomic volume and atomic potential energy of the atoms at and near the Fe-Cu interface supports the WS identification of point defects there.

Fig. 4 shows surviving defects versus PKA energy. In these simulations vacancies mostly remain at the Fe core for both NP and CSNP, but interstitials are significantly depleted in the CSNP Fe core. Many defects relocate to the surface of the NP and CSNP. Defect migration to the surface is a well known healing mechanism present in nanofoams [9,12,31,78,79]. The number of vacancies in bulk Fe simulations is similar to the number in the Fe core of CSNP. On the other hand, the number of interstitials is much lower than for bulk cascades. This behavior agrees with experimental findings where surface boundaries play a relevant role in high surface-to-volume ratio nanostructures [11,28]. We note that the ability of the FeCu interface to store both vacancies and interstitials grows faster with PKA energy than defect storage in the core, leading to a cross-over of the corresponding lines in Fig. 4. The number of surviving defects in the NP is higher than in bulk, from our results in bulk iron and results obtained by Björkas [77], Granberg et al. [23] and Bacon et al. [14] using similar EPKA and different interatomic potentials. This is partly because the FP count for the NP includes atoms relocalized to the surface.

For the CSNP cases, the presence of the Cu shell minimizes the accumulation of vacancies in the interface and to a lesser extent, also in the core. The total vacancy count on each type of NP is similar, but the CSNP concentrates these defects in the Cu shell. Despite this, we do not observe significant vacancy clustering in the shell during primary damage. Interstitials follow a similar trend than vacancies: although there are almost no differences in the total number of interstitials produced in the NP or CSNP, approximately 75%-80% of them are stored in the Cu shell as Cu interstitials for the CSNP case.

For Fe, single vacancy migration energy is ~0.6 eV, while the dumbbell migration energy is ~0.3 eV [80]. The vacancy diffusivity is ~$10^{-16}$ m$^2$/s at room temperature [81], and this means that it will take ~0.001 s for single vacancies in the Fe core to reach the NP surface. Di-vacancy diffusion has similar activation energy [82]. Therefore, primary damage will likely evolve and mostly disappear due to defect migration to the surface of the NP and CSNP during experimental time scales, for moderate dose rates.

Recombination rates ($\eta$) are presented in Fig. 5, including only vacancies. Recombination is higher than in bulk Fe [18] with different potentials and similar to our bulk values. The lower recombination rate of NP compared to CSNP is partly due to the large Fe sputtering yield, but the rate is still large due to the surface acting as a defect sink. The CSNP displays a higher recombination rate, with improved recovery from primary damage, This can be associated with larger radiation resistance, and occurs thanks to the Cu shell, and might be akin to the experimentally observed effect of an oxide layer reducing defects on irradiated steel NP [42].

Fig. 6 shows the comparison of MD with the arc-dpa model developed by Nordlund and coworkers [58]. The pure Fe NP follows the same trend for vacancies and interstitials with an imbalance between them because of the sputtering yield. For the CSNP, there is a large imbalance between interstitials and vacancies due to interstitial migration to the CS interface. Both the CS interface and the Cu surface act as defect sinks. Interstitial mobility is higher than vacancy mobility [83], explaining the remnant vacancies in the Fe core, with reduced clustering. Finally, our bulk simulations show a reasonable agreement with the analytical arc-dpa model and also with the presence of vacancies in the CSNP. In order to check the possible influence of our ZBL matching procedure in the results [85,86], we have also carried out a few simulations with a slightly different ZBL fit by Zhang et al. [22], to check for possible differences in defect production. Damage was found to be the same for both potentials.

In our simulations, radiation damage is greatly influenced by sputtering from the NP, which is given in Fig. 7. Panel 7(a) shows the sputtering yield as a function of the PKA energy for both the Fe NP and the FeCu CSNP. The latter is a consequence of the eyecta proceeding from the core ($CSNP_{Fe}$) and the shell ($CSNP_{Cu}$). In this case, the total quantity almost matches the ejecta that originates in the shell, and the difference that corresponds to the Fe sputtering yield falls at least one order of magnitude when compared with the Fe NP case. Since our PKA model always starts from a random Fe particle in the core, it is plausible that the shell provides an effective structure for trapping the displacing particles trying to leave the CSNP, minimizing the ejections. Panel 7(b) displays sputtering time evolution for a PKA energy of 3 keV, showing that ejection ends after 4 ps. Similar times are obtained for the other PKA energies in this study. Available experimental data for Fe on Fe irradiation from Eckstein [70] gives $Y$=3.2 from a flat surface, at normal incidence, in good agreement with our calculations using BCA for a thin film giving $Y$=3.9 and shown in Fig. S7.

The high surface-to-volume ratio in many nanomaterials favors enhanced sputtering yields, $Y$, for instance in nanofoams [78,79]. Zimmerman et al. [35] found, for Au NP bombarded with energies between 16 and 64 keV, a more than twice higher sputtering yield when compared with planar targets. Urbassek et al. [36] found enhanced sputtering yield when the nanoparticle size is of the order of the projectile range. Jurac et al. [34] used MC to simulate sputtering from spherical carbon particles bombarded by He and H ions, and found significant enhancement compared to a flat surface only for $R/R_p$<1 (where R denotes NP radius). The enhancement of $Y$ in all those cases is related to ejection from the side and back surfaces by the particle initiating the collision cascade and the recoils. Sputtering increases with PKA energy, unlike what was observed for a W NP [65], showing a maximum below 5 keV. For our CSNP, $D/R_p$~2.6, using $R_p$=3.4±1.7 nm from *SRIM* for 5 keV Fe on bulk Fe, and low NP enhancement would be expected based on the work by Jurac et al. [34]. However, the enhancement factor with respect to a thin film simulated with *SRIM* is still large because we are impacting a metal NP with a PKA, instead of bombarding C with light ions, as in [34]. For the Fe NP, sputtering at 1 keV has an enhancement factor ~2, while at 7 keV it is ~10.

We note that sputtering reduction from the Fe core could be tailored, both changing the shell thickness and composition. For instance, a Ni shell of the same thickness might reduce sputtering further, given that Ni has a larger binding energy than Cu. One could consider a collection of FeCu CSNP, forming a film that would protect some target. In this case, another effect has to be taken into account, because ejecta from one particle can cause re-sputtering from the substrate and from other NP. This is akin to what happens in the sputtering of foams, as simulated by Rodriguez-Nieva for swift heavy ions [78] and by Anders et al. for keV projectiles impacting Au nanofoams [79]. On the other hand, the cavities in a NP film might also store the low energy ejecta and provide surfaces as effective defect sinks after long-time defect evolution. For a NP array, redeposition of sputtered materials can also change the NP size [38,86,88].

## 3.1. Cumulative bombardment

As proof of principle, we carried out some cumulative bombardment simulations for both nanoparticles. Fig. 8 shows snapshots of the cumulative 3 keV simulation for FeCu. Each individual PKA generates around 240 displacements according to the RPA model [58] (see Fig. S8). Therefore, all the 50 simulated PKA would generate a cumulative dose of ~0.81 dpa for the Fe NP and ~0.40 dpa for the CSNP. For bulk, the number of defects increases with dose and there is formation of dislocation loops [89]. We also observe an increase in the number of defects in our NP simulations, with some more clustering, but without any dislocation loops in the Fe core. For our 3 keV simulations a linear extrapolation of the single PKA results would give ~550 vacancies and 65 interstitials in the Fe core. We found 67 vacancies and 3 interstitials dumbbells in the Fe core after 50 PKA for the CSNP, and 72 vacancies and zero instertitials in Fe core for the NP. The larger vacancy cluster for the first case contains 13 vacancies and 37 for the second sample. These numbers indicate that defect sinks at the surface and the interface, remain active and show no saturation even close to a 1 dpa. With regards to ejections, we found 280 Cu and 17 Fe atoms sputtered from the CSNP, while the NP produced 911. Fe sputtering is larger than a linear extrapolation from individual PKA results because the Cu shell displays some porosity at the end of the cumulative bombardment, as can be seen in Fig. 8. Radiation resistance at higher doses will require thicker shells than the one considered here, which is only 1 nm thick.

We also evaluated the energetics of ejecta as a potential cause of further damage caused by re-sputtering impacting adjacent nanoparticles in a nanostructured material integrated by a collection of NP. Fig. S9 shows a normalized histogram for the kinetic energy spectra of the sputtering yield. Both systems behave with the expected $E^{-2}$ decay, in reasonable agreement with the analytic model by Sigmund-Thompson for binary collisions [90]. This fact, along with the relatively low sputtering shown in Fig. 7, indicates that re-sputtering from ejecta will be negligible since most ejecta will re-deposit. The net decrease of total sputtering yield due to redeposition of ejecta has been recently studied for cumulative bombardment of a flat W surface with a dome, equivalent to a half-buried nanoparticle [91]. Further discussion on sputtering is available in the Supplementary Material.

## 4. Conclusions

There is a need for radiation-resistant nanostructured materials [10,92]. In this work we focus on the irradiation of nanoparticles (NP), including a Fe NP and a FeCu core-shell NP (CSNP), which has been studied using Molecular Dynamics (MD) simulations, aiming to uncover possible radiation resistance. NP with a Fe core, and a Cu shell can be produced in the lab and atomistic simulations can provide a reasonable description of those NP [46]. In this study, the Fe core has ~7 nm diameter. Using PKA energies up to 7 keV produces point defects and a few small defect clusters, similar to what happens in bulk [23]. Point defects in the NP appear as interstitial dumbells and vacancies. No twins were observed in the Fe core, but there are some pre-existing stacking faults crossing the Cu shell and they can be modified by the irradiation. There is no significant rugosity nor volume increase in the NP due to the irradiation. The projectile range is about ⅓ of the NP diameter for the 5 keV PKA, but there is significant Fe sputtering from the side and back surfaces, much larger than that from a flat surface or from a thin film. This large sputtering decreases defect recombination for the NP. Khan *et al.* [43] simulated 15 nm Fe NP bombardment and found very few surviving FP, but some large defect clusters and less sputtering, as expected from a larger NP.

We find that covering the Fe NP with a Cu shell reduces significantly the number of surviving interstitials in the Fe core, while also diminishing the Fe sputtering yield, and increases defect recombination in the Fe core. This is in qualitative agreement with experiments CSNP, where it was suggested that the CSNP ability to prevent sputtering and accommodate defects resulted in the observed radiation resistance [49,50].

MD results are contrasted to a Monte Carlo BCA simulation with SRIM2016 [66] for equivalent thin films [48-50], and also to the arc-dpa model [58] for bulk Fe. The Fe NP displays large sputtering yield, leading to a large number of defects, compared to bulk. The CSNP shows large Cu sputtering and a very defective Cu shell. Radiation damage in the Fe core of the CSNP is quite different from bulk results: the number of vacancies is similar to predictions from arc-dpa, while there is a very low number of interstitials in the Fe core, because the Fe-Cu interface and the Cu shell act as defect sinks. Therefore, our results indicate enhanced radiation stability for the CSNP.

We simulate primary damage, but defect evolution during experimentally relevant times has to be modeled using alternative methods, such as kinetic Monte Carlo [93,94], but appropriate inclusion of surface and interface sinks remains challenging. The Cu-Fe interface and the NP surface are expected to act as a defect sink and further reduce the defect content, improving NP radiation resilience. Preliminary results from annealing simulations show this recovery.

Radiation can change the mechanical and chemical properties of core-shell nanoparticles, possibly enabling additional technological applications [95], like improved hardness or enhanced chemical reactivity. Cumulative irradiation leads to differences in defect storage and sputtering [26], compared to the single PKA irradiation. There are no dislocations formed in the Fe core of the CSNP and low vacancy storage compared to bulk. Bombardment at higher energies, where the collision cascade becomes highly non-linear and thermal spikes can dominate [36,37], will be required to assess the possible application in nuclear environments, where PKA energies of tens of keV might destroy or fragment NP [96]. However, astrophysical shielding applications from keV solar wind seem feasible [7,8].

All in all, we focus on the effects of irradiation in a single CSNP case with a fixed core size and shell thickness. However, NPs size effects deserve to be further studied in order to reveal the transition from point defect to dislocation loop PKA induced damage [43]. Besides, both parameters, internal radius and shell thickness, or even nanoparticle shape, can be easily controlled in order to tailor the radiation response of finite size nanostructures. Nanoparticles present advantage in regard to synthesis methods, allowing a completely controllable manufacture in composition, shape and size, for almost every element in the periodic table [45]. We have studied nanoparticles with a single component metallic core. Other nanoparticles might display more complex behavior, including amorphization [97]. The possibility of radiation mitigation in CSNP together with synthesis advantages opens a wide scenario to obtain novel radiation resistance materials, like thin films composed of NPs. Tools from machine learning and material databases might help in designing enhanced radiation resistant NP in the near future [98-100].

**CRediT authorship contribution statement**

D.R. Tramontina: Software, Formal analysis, Writing- Original Draft, Visualization, O.R. Deluigi: Software, Formal analysis, Visualization, Data curation, Writing - Review & Editing, Visualization. R. Pinzón: Methodology, Software, Formal analysis, Resources, Funding Acquisition. J. Rojas-Nunez: Software, Methodology, Writing - Review & Editing. F.J. Valencia: Software, Formal analysis, Writing - Review & Editing. R.C. Pasianot: Software, Methodology, Writing - Review & Editing. S. E. Baltazar: Methodology, Writing - Review & Editing, R. I. Gonzalez: Methodology, Writing - Review & Editing, E.M. Bringa: Conceptualization, Methodology, Formal analysis, Writing- Original Draft, Writing - Review & Editing, Funding Acquisition, Supervision.

**Acknowledgements**

We thank D. Schwen and X. Bai for kindly providing their interatomic potential tables. DT acknowledges support from DIUM. DT, OD and EMB thanks support from SIIP-UNCUYO 06/M008-T1 and PICTO-UM-2019-00048. We used Serafin Cluster (CCAD-UNC) and Toko (UNCuyo) clusters, which are part of SNCAD-MinCyT, Argentina. RP acknowledges a scientific fund from Sistema Nacional de Investigación de Panamá (SNI) and SENACYT Projects: FID-2016-275 and EIE18-16. HPC-Cluster Iberogun Group gratefully acknowledges the support of NVIDIA Corporation with the donation of the Titan Xp GPU used for this research. FV an RIG thank Fondo Nacional de Investigaciones Científicas y Tecnológicas (FONDECYT, Chile) under grants #1190662, #11190484 and #11180557. This research was partially supported by the supercomputing infrastructure of the NLHPC (ECM-02). SEB, FV, and JRN acknowledge the support from the Basal Program for Centers of Excellence, Grant AFB180001 CEDENNA, CONICYT. RCP acknowledges support from project PICT 2019-02912 ANPCyT.

**Figures**

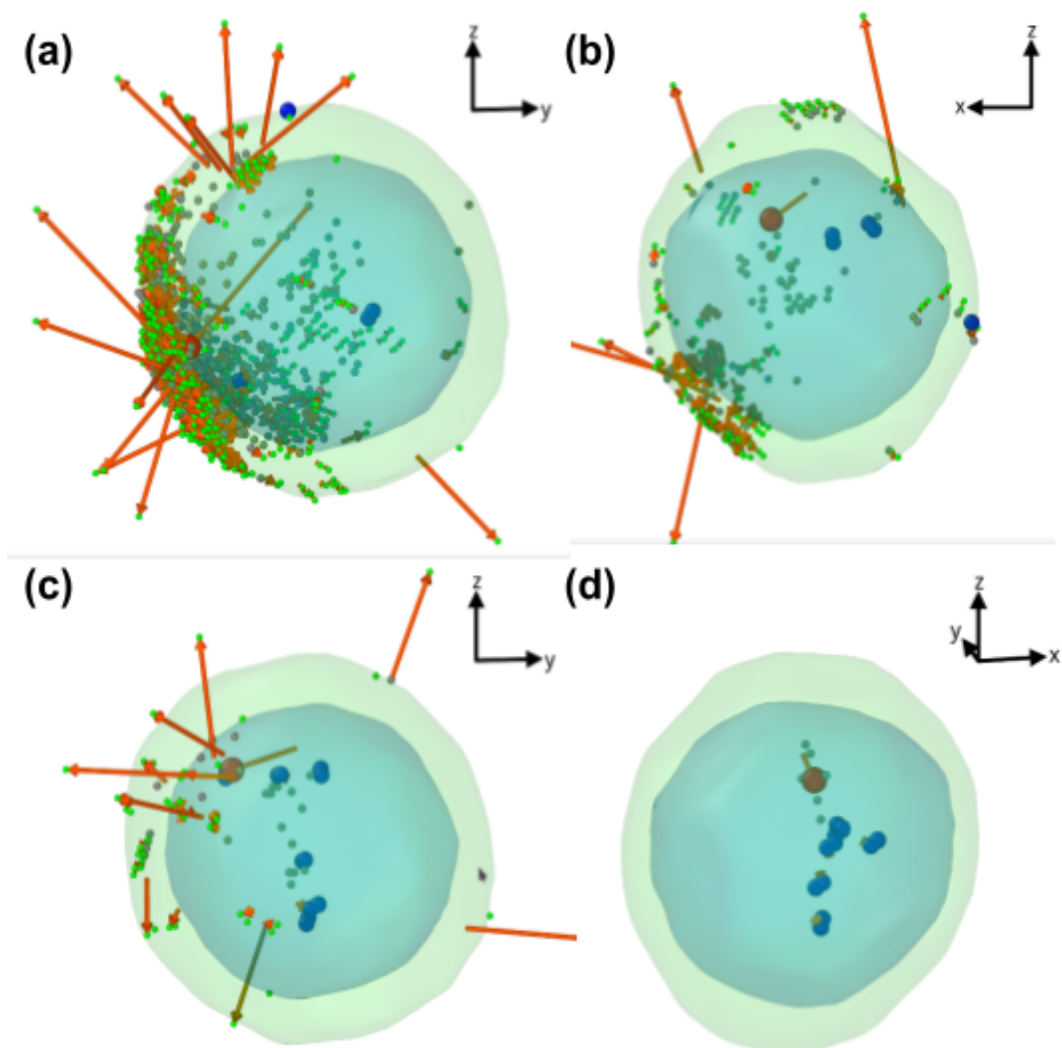

**Fig. 1.** Cascades produced with the same pka incident direction at several initial energies, for the CSNP. Snapshots taken between 5 ps and 7 ps, after most of the recombination occurs. Arrows indicate the trajectory of the displaced atoms. (a)-(d): 7, 5, 3 and 1 keV. Translucid surfaces indicate the shell and core surfaces. Only defective atoms are shown: Fe PKA (red); Fe atoms at interstitial positions (blue); sputtered and interstitial Cu atoms as well as Cu adatoms in the CSNP surface (green); vacancy sites (gray). Fe defects and the PKA are displayed with larger size. Arrows (orange) go from initial to final position of defective atoms. Note that Fe interstitials are dumbbells.

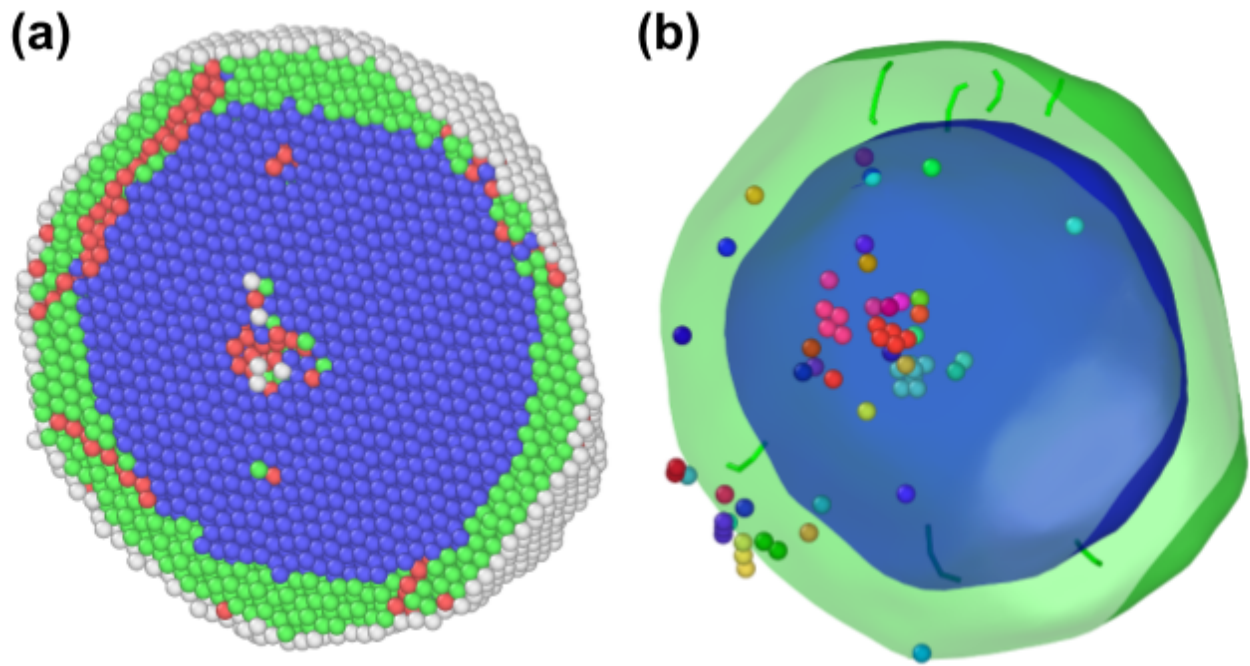

**Fig. 2.** Snapshots after defect recombination for 5 keV, at~ 9.8 ps. (a) Only half the NP is shown, slicing with a (100) plane. (b) Only half of the NP surfaces are shown, but all defective atoms are included. Pannel (a): atoms colored by PTM structure: bcc (blue); fcc (green); hcp (red, stacking faults), other structures, including surface atoms (white). Panel (b): vacancy sites colored by cluster (red largest cluster); core-shell interface (blue surface), NP surface (green surface). Preexistent Shockley partials are detected by DXA on the shell and displayed as green lines. No dislocations are detected in the bcc Fe core, which only contains small point defect clusters.

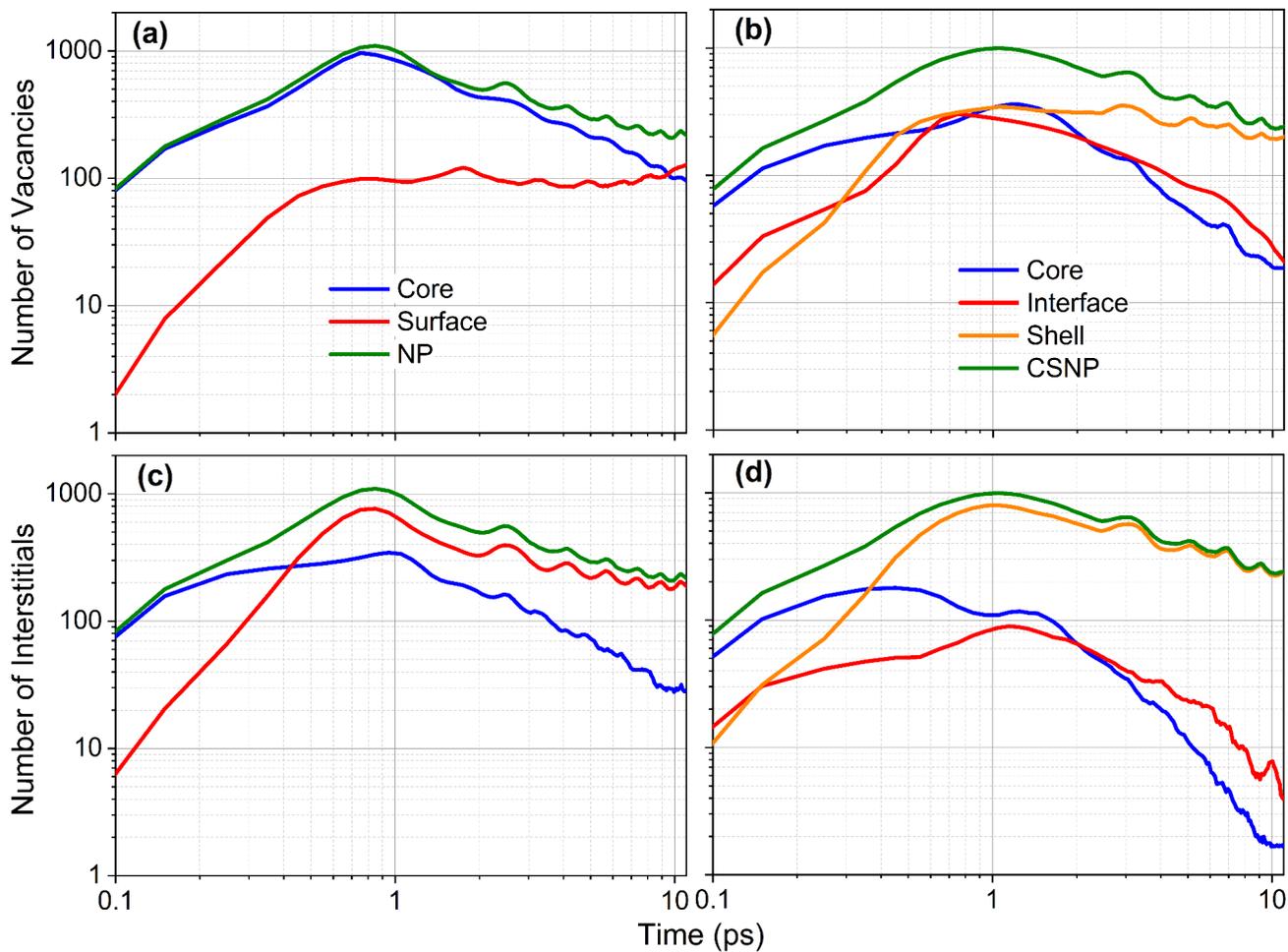

**Fig. 3.** Damage evolution for 5 keV PKA. (a) Vacancy count for Fe NP and (b) FeCu CSNP. (c) Interstitial count for Fe NP and (d) FeCu CSNP. Damage at the FeCu interface and core is very small after a few ps.

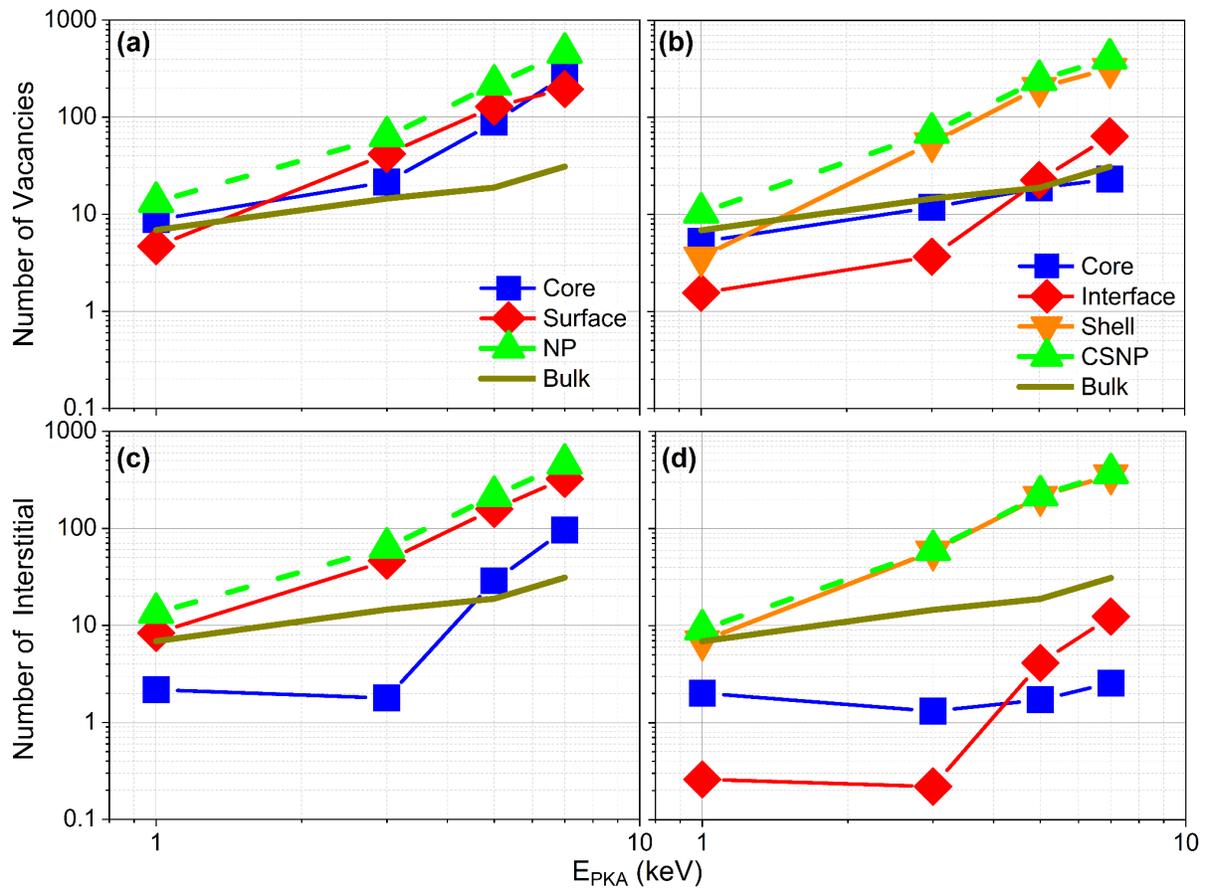

**Fig. 4.** Surviving defects evaluated at 11 ps, at the end of the cascades. Vacancy counts for: (a) Fe NP, and (b) FeCu CSNP. Interstitial count for: (c) Fe NP and (d) FeCu CSNP. Also we compare all cases with bulk Fe. The Cu shell lowers the number of remaining defects at the Fe core, when compared to the pure Fe NP case.

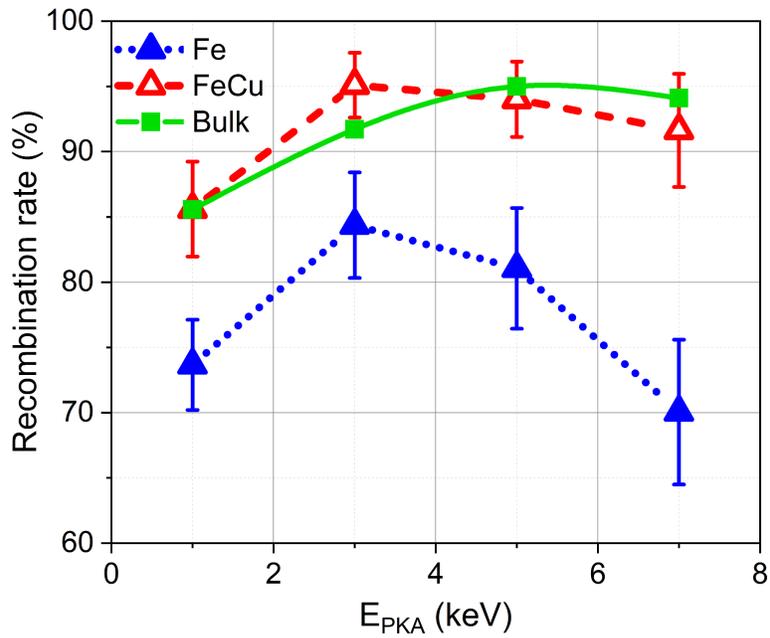

**Fig. 5.** Recombination rate as a function of the PKA energy. Here we consider vacancy recombination and do not include Cu vacancies from the shell of the CSNP. As expected, recombination in the CSNP is higher than in the NP, but slightly lower than in bulk.

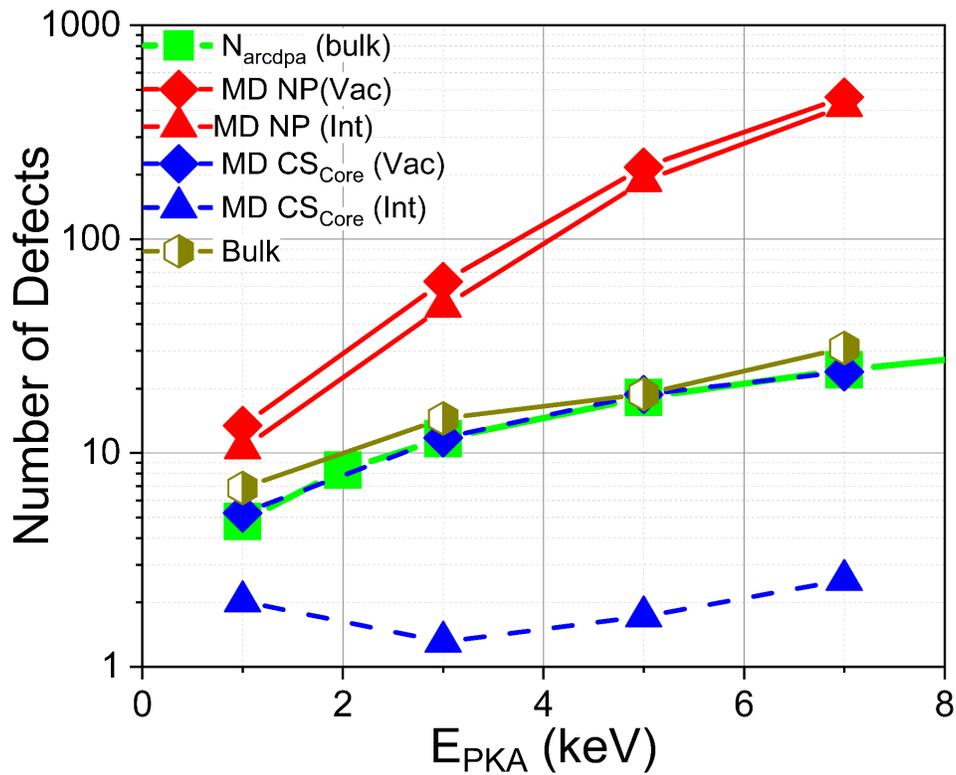

**Fig. 6.** Comparison between MD results for NP, CSNP and bulk samples, and the arc-dpa model [58].

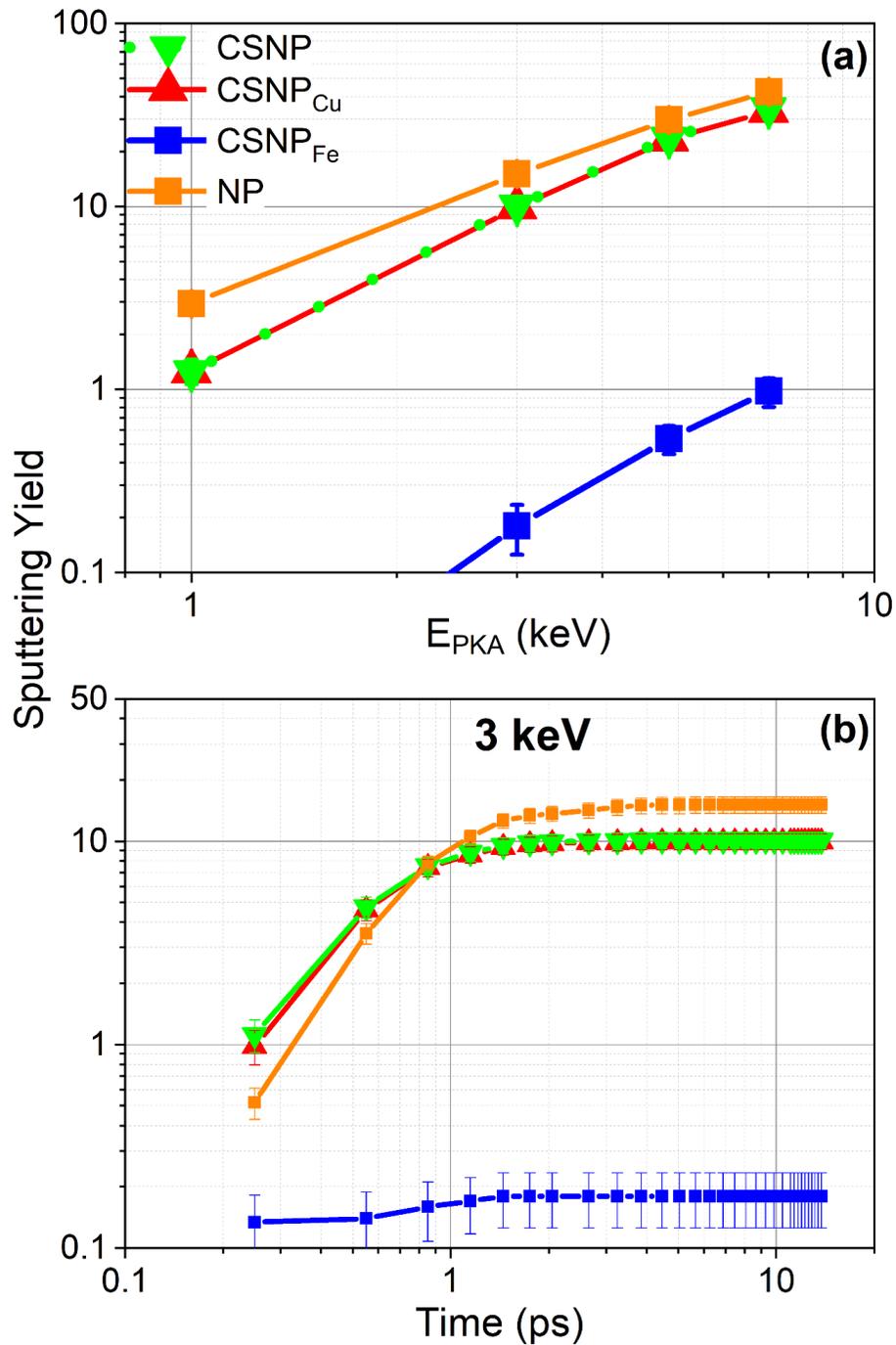

**Fig. 7.** (a) Sputtering yield for the Fe NP (MD), FeCu CSNP (MD), and the FeCu thin film (SRIM) as a function of radiation energy. Contributions from core and shell are also computed for the CSNP. The PKA is included in these calculations, when sputtered. Values are evaluated at 11 ps. Sputtering is lower for the CSNP, and originates mostly from the Cu shell, lowering the contribution from the core compared with the single Fe NPe. (b) Evolution of the sputtering yield for 3 keV cases, showing that sputtering reaches a final value after a few ps.

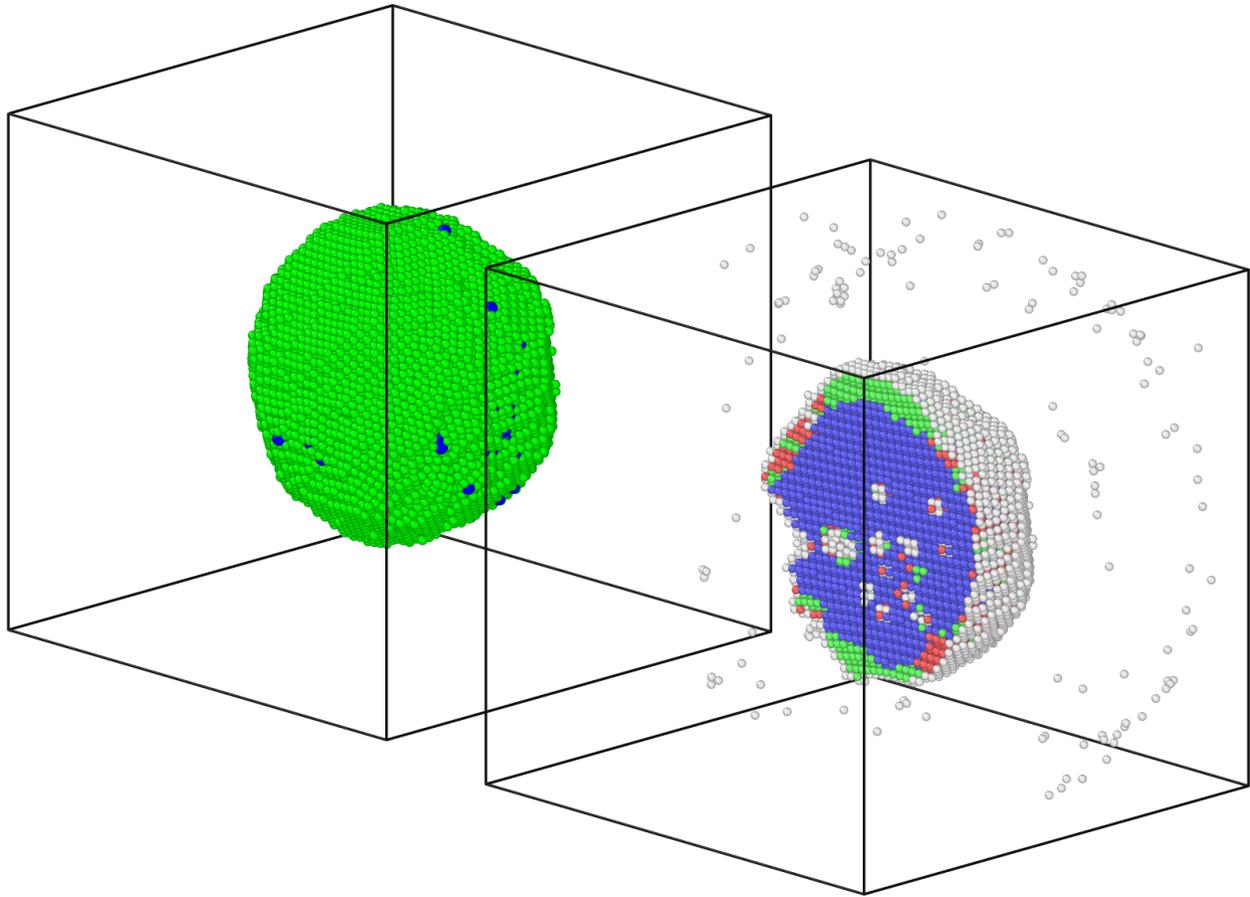

**Fig. 8.** Final snapshot for cumulative bombardment with 50 PKA of 3 keV for the CSNP. Left panel: Fe atoms (blue) and Cu atoms (green), indicating some shell porosity after ~0.8 dpa damage. Right panel: Only half the NP is shown, slicing with a (100) plane. Atoms colored by PTM structure: bcc (blue); fcc (green); hcp (red, stacking faults), other structures, including surface atoms and sputtered atoms (white). Most of the sputtering yield proceeds from the Cu shell and a few Fe particles appear retained in the CSNP surface.

# Supplementary Material

The SM includes Figs. S1-S9, and a description of related Monte Carlo simulations.

## Monte Carlo (MC) Simulations

**I-MC Methods**. For a simple comparison with MD, the *SRIM2016* package was used to perform the Monte Carlo (MC) simulations [1] employing Fe thin films (TF) with thickness similar to NP and CSNP diameters, and a three layer Cu-Fe-Cu target where both front and back depths equaled the shell NP Cu thickness. This approach was already used to help interpret experiments on irradiation of CSNP [2-4]. There are BCA codes that allow for the modeling of crystals, i.e. *Marlowe* [5] or *CrystalTrim* [6], and codes like *TRI3DYN* which also allow modeling of nanostructures [7]. Here, surface binding energy ($U_0$) was 4.34 eV for Fe as suggested by *SRIM* and also used by Eckstein [8], whereas for Cu $U_0$=3.52 eV. Displacement energies were 40 eV for Fe and 33 eV for Cu [9]. Although BCA does not allow for defect recombination, and the amorphous media will not provide a reasonable structure for the Fe-Cu interface, nor for complex surface binding, this technique will allow for the simulation of an immense number of trajectories with improved statistical results. In this case, the infinite lateral extension of the TF will also modify defect evolution compared to the NP. Using SRIM2016 we calculated the penetration for the *5* keV projectiles into the Fe film, given by the projected range ($R_p$) [10], obtaining $R_p$=3.2nm, with longitudinal straggling $s_{//}$=2.0 nm, and lateral straggling $s_\perp$=1.3 nm. MC simulations of cascades used $25 \times 10^4$ projectiles, ensuring that results converged to steady state values.

**II-MC results:** Fig. S5 Show a normalized histogram for the kinetic energy spectra of sputtered atoms as determined by MD simulations for 5 and 7 keV PKA for both Fe and FeCu NP. There is reasonable agreement with the analytical expression from Sigmund-Thompson for binary collision cascades [11], indicating a $E^{-2}$ decay. Note that the highest energy simulated in our work produced at most, some resputtering at ~ 4 keV, while 5keV PKA produces only some 3 keV ejections, whose resputtering is negligible.

Fig. S6 is similar to Fig. 6 but adding MC results for a 7 nm Fe film simulated with *SRIM*. The NP shows values close to *SRIM* at 1 and 3 keV, but much larger than *SRIM* for larger energies. This is expected, given that *SRIM* models a film instead of a NP. The number of vacancies and interstitials are similar and follow the same trend with energy, as expected. *SRIM* does not include defect recombination, and would give larger values than bulk simulations. A recent BCA MC model [12], coupled with the Athermal Recombination Corrected Displacements Per Atom (arc-dpa) model [13], would give a more accurate estimation of bulk damage. arc-dpa is fit to bulk cascades, but provides defect numbers which are extremely close to the number of vacancies in the Fe core of the CSNP. However, the amount of interstitials present in the Fe core is much less than the amount predicted by this analytical bulk model, explaining the lack of interstitial clusters and dislocation loops.

Fig. S5, is also similar to Fig. 7(a) in the main text, adding MC results. The BCA simulations do show some transmitted projectiles but no sputtering from the back surface. We note our PKA simulations have a range of angles, which would enhance sputtering from a flat surface. In addition we also have ejection from side surfaces, enhancing *Y*. BCA with *SRIM* uses an amorphous target, without partial channeling which would enhance back surface sputtering, as in MD.

For the CSNP, sputtered atoms are almost entirely Cu atoms, and this agrees with *SRIM*, as observed in Fig. S4. The Cu shell acts as an effective barrier for Fe ejection. We note that in *SRIM* sputtering occurs only from the top surface, while for the NP and CSNP there is sputtering mostly from the upper hemisphere of the NP surface. We have also considered a pure Fe thin film, 9 nm thick. Sputtering decrease is negligible for this size increase, as shown in Fig. S4. Despite this lack of variation for a thin film, a larger NP might lead to reduced sputtering from side surfaces.

**Supplementary Figures**

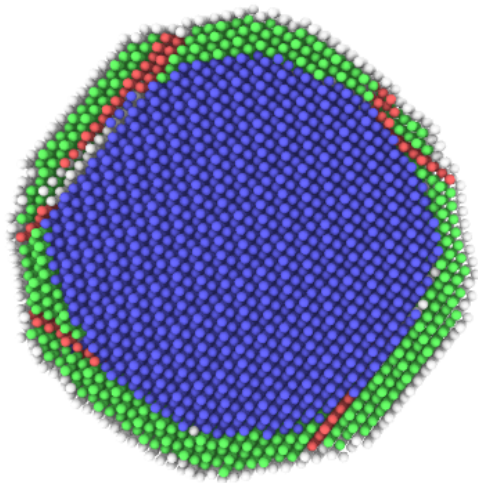

**Fig. S1:** CSNP before irradiation. The Fe core is a perfect bcc crystal (blue). The fcc Cu shell (green) includes some stacking faults, identified as hcp atoms (red). Some atoms with unidentified phase (gray) appear at the FeCu interface, and at the surface.

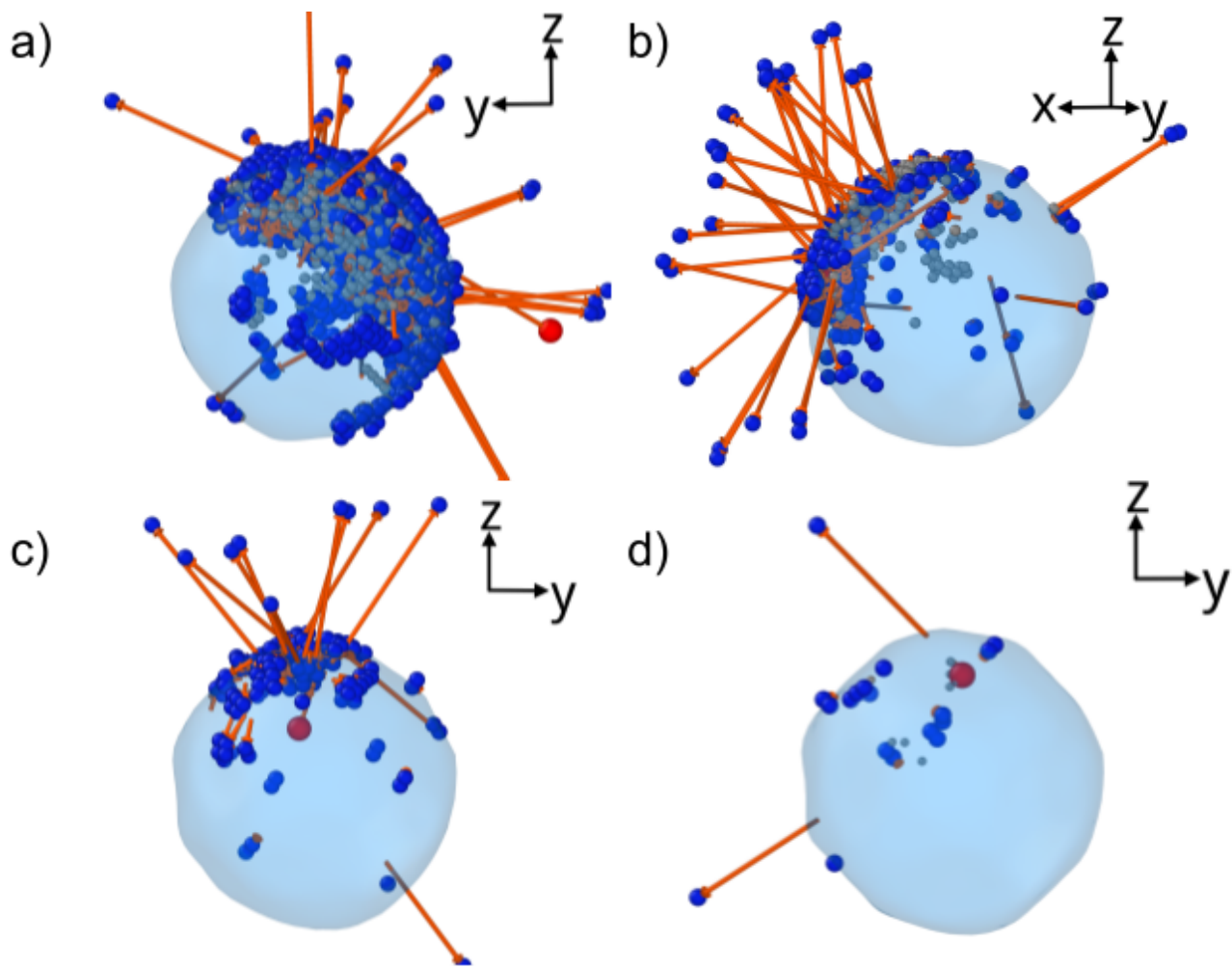

**Fig. S2:** Cascades produced with the same pka incident direction at several initial energies, for the Pure Fe nanoparticle, 9 ps after PKA event. (a)-(d): 7, 5, 3 and 1 keV. Translucid surfaces indicate the core surfaces. Only defective atoms are shown: Fe PKA (red); Fe atoms at interstitial positions (blue); ; vacancy sites (gray). Fe defects and the PKA are displayed with larger size. Arrows (orange) go from initial to final position of defective atoms. Note that Fe interstitials are dumbbells.

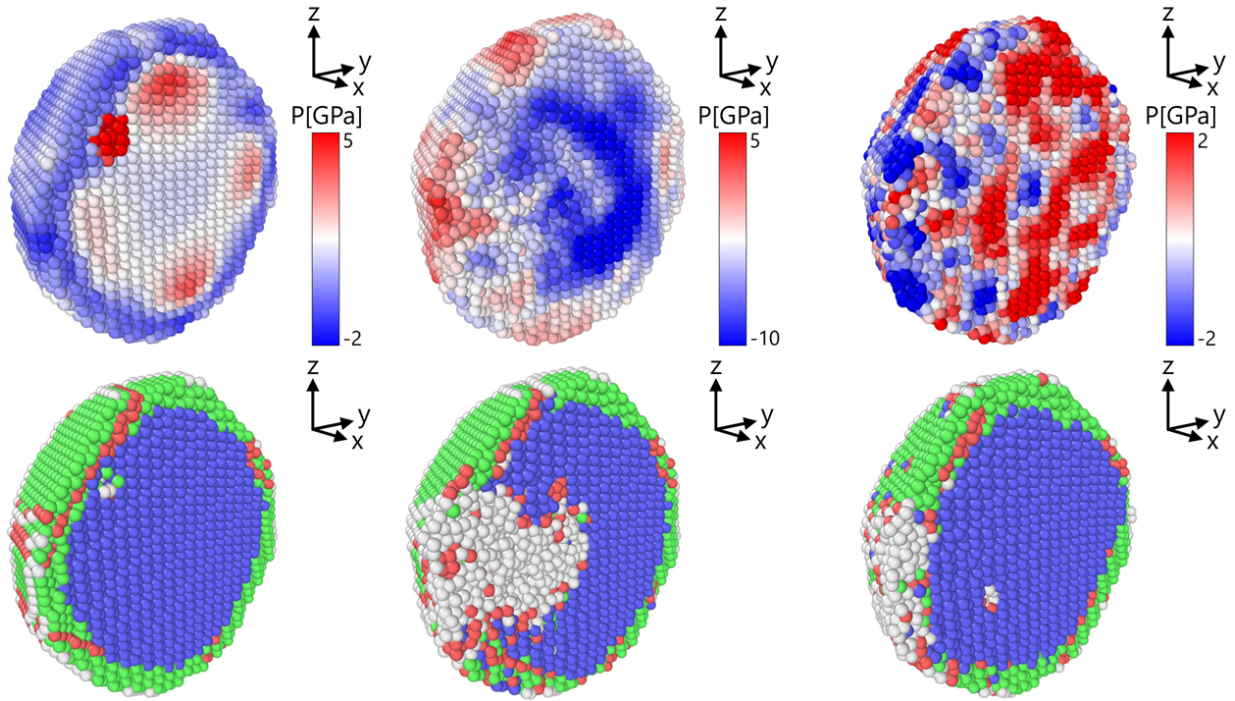

**Fig. S3:** Snapshots of a typical 7 keV PKA in the CSNP, for different times after the PKA. Left: ~0.01 ps, middle: ~1 ps; right: ~11 ps. Top row: coloring with atomic stress, calculated using Voronoi atomic volumes, and averaging over a sphere with radius 0.5 nm for each atom. Atoms with Voronoi volume larger than 0.020 nm$^3$ (surface atoms) were removed. Bottom row: coloring by PTM, with rmsd 0.02nm. The Fe core is a perfect bcc crystal (blue). The fcc Cu shell (green) includes some stacking faults, identified as hcp atoms (red). Some atoms with unidentified phase (gray) also appear. In the earlier frame, the PKA appears in the center of the bright red cluster (top row), and as one of the gray atoms in the bottom row. The initial tensile stress at the Cu shell turns into highly compressive stress of few GPa at 1 ps, which leads to creation/removal of stacking fault atoms near the bottom/top of the CSNP in the final frame. Stress in the core will influence interstitial migration.

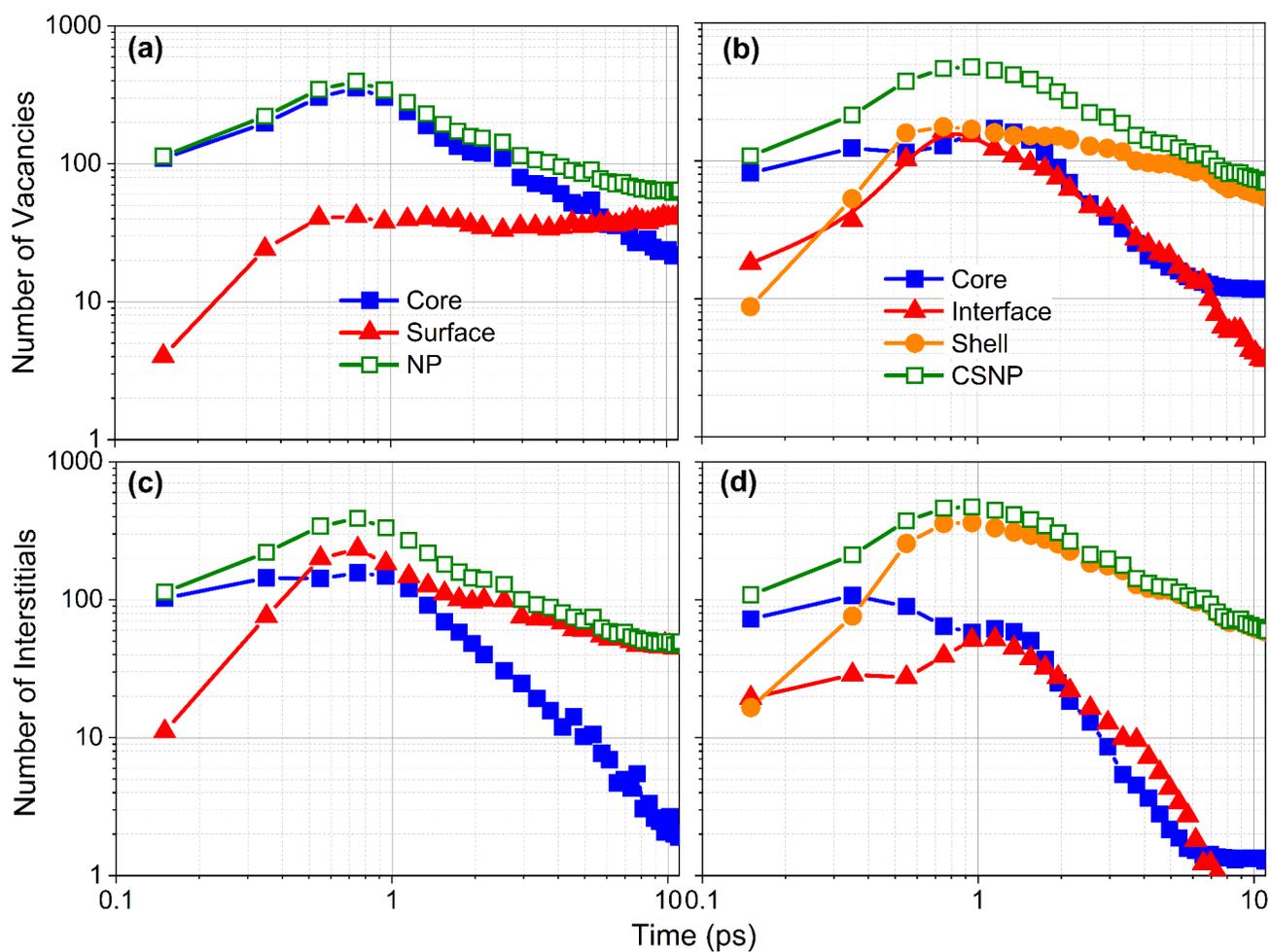

**Fig. S4:** Damage evolution for 3 keV PKA. (a) Vacancy count for Fe NP and (b) FeCu CSNP. (c) Interstitial count for Fe NP and (d) FeCu CSNP. Damage at the FeCu interface and core is very small after a few ps.

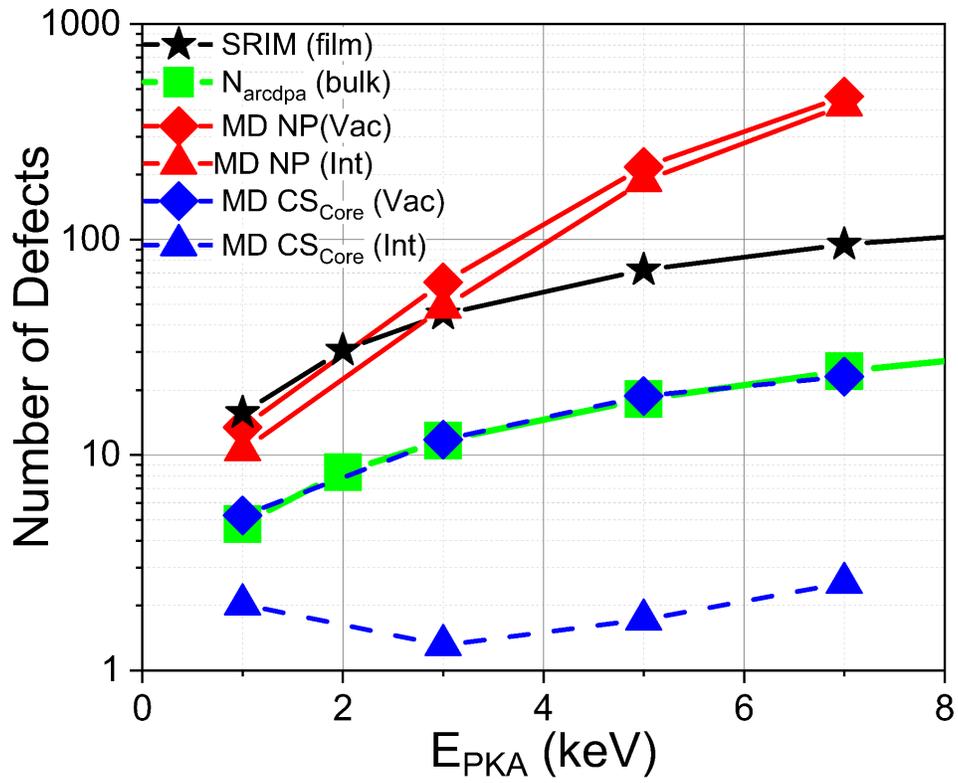

**Fig. S5:** Comparison between MD results and models. SRIM for 7 nm Fe film.

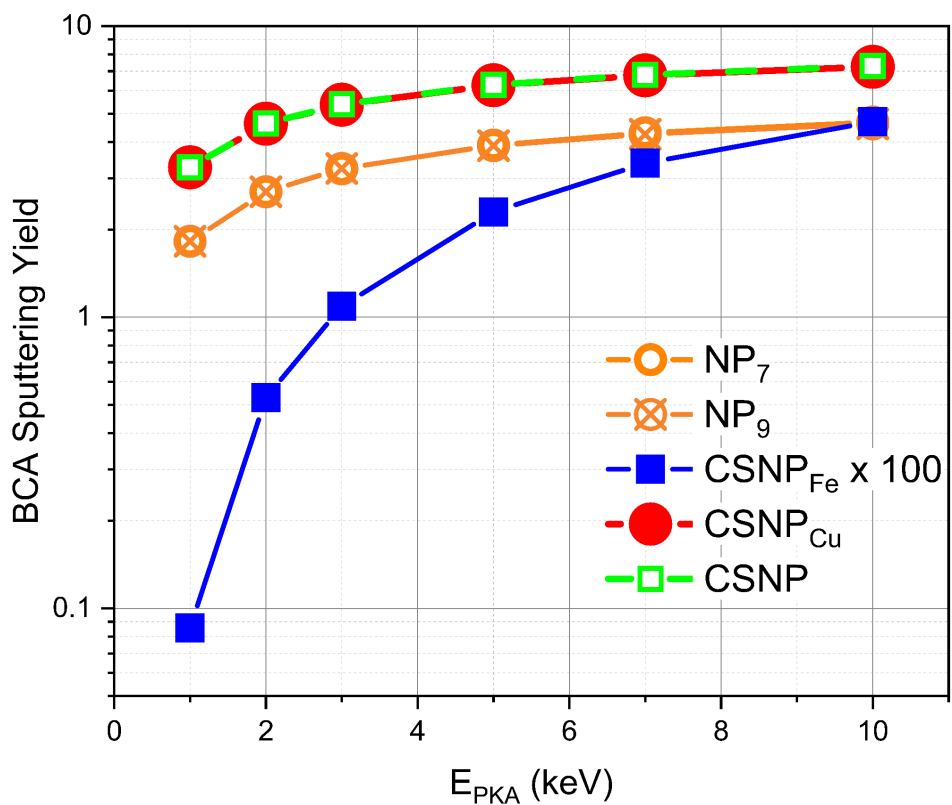

**Fig. S6:** BCA Monte Carlo simulations of sputtering yield using *SRIM* vs. MD results. For MC calculations, two thicknesses were evaluated for the pure Fe case: 7 nm ($NP_7$) and 9 nm ($NP_9$). The Fe NP simulated with MD has a 7 nm diameter. The FeCu film has a 7 nm Fe central layer recovered by 1 nm Cu on each side, for a total thickness of 9 nm (CSNP), analogous to the CSNP. Sputtering yield is similar for $NP_7$ and $NP_9$ TFs as expected, since it originates from the upper surface. Cu sputtering from the CuFeCu film, emulating CSNP, is larger because it consists mostly of Cu ejecta. Note that Fe ejection is more than two orders of magnitude lower than Cu ejection.

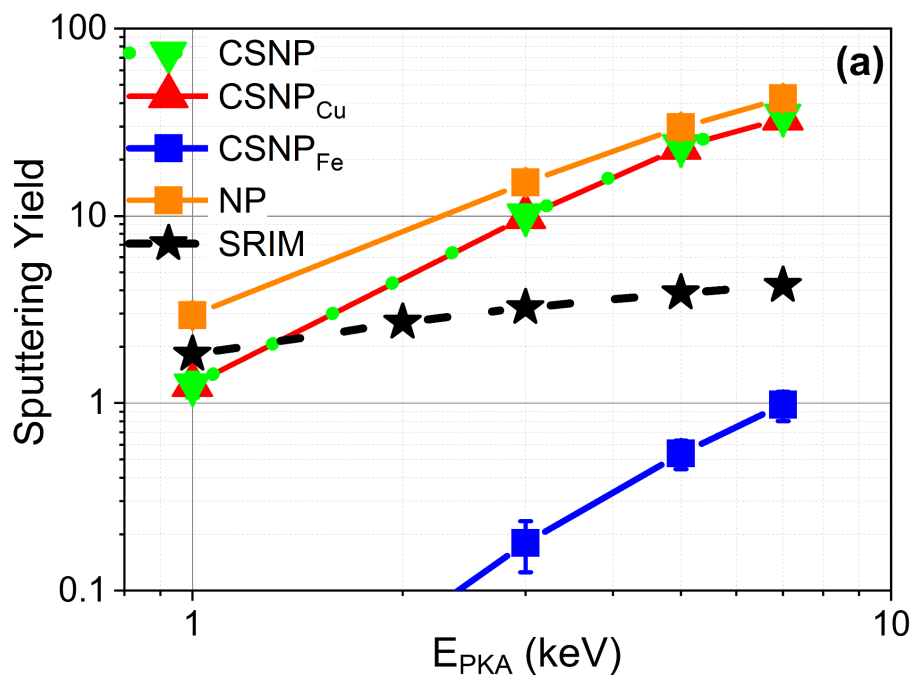

**Fig. S7:** Sputtering yield for the Fe NP (MD), FeCu CSNP (MD), and the FeCu thin film, 7 nm thick (SRIM) as a function of radiation energy. Contributions from core and shell are also computed for the CSNP. The PKA is included in these calculations, when sputtered. Values are evaluated at 11 ps. Sputtering is lower for the CSNP, and originates mostly from the Cu shell, lowering the contribution from the core compared with the single Fe NPe.

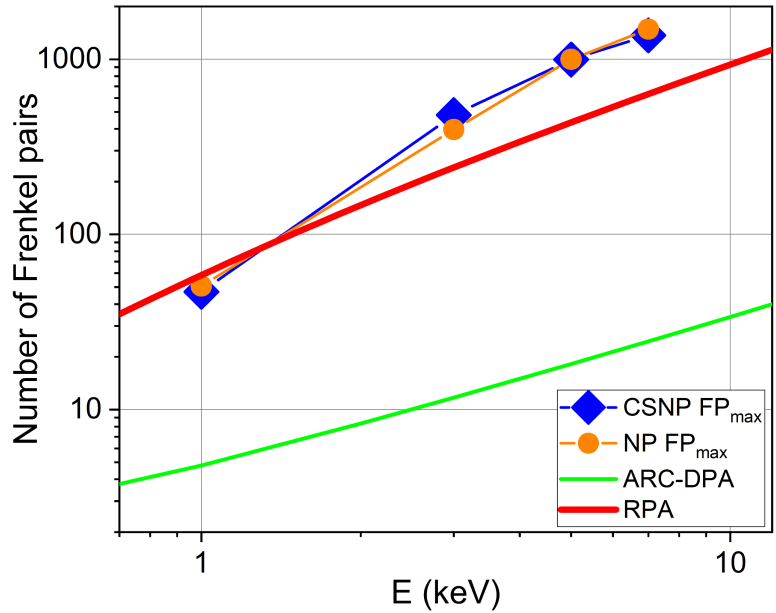

**Fig. S8:** Number of Frenkel pairs: peak number of FP from MD simulations, $FP_{max}$, arc-dpa FP, and the corrected number of displacements, given by the "Replacements Per Atom" model, RPA [9]. Note that $FP_{max}$/RPA~2 above 1 keV. RPA and ARC-DPA were calculated with Fe parameters from [9].

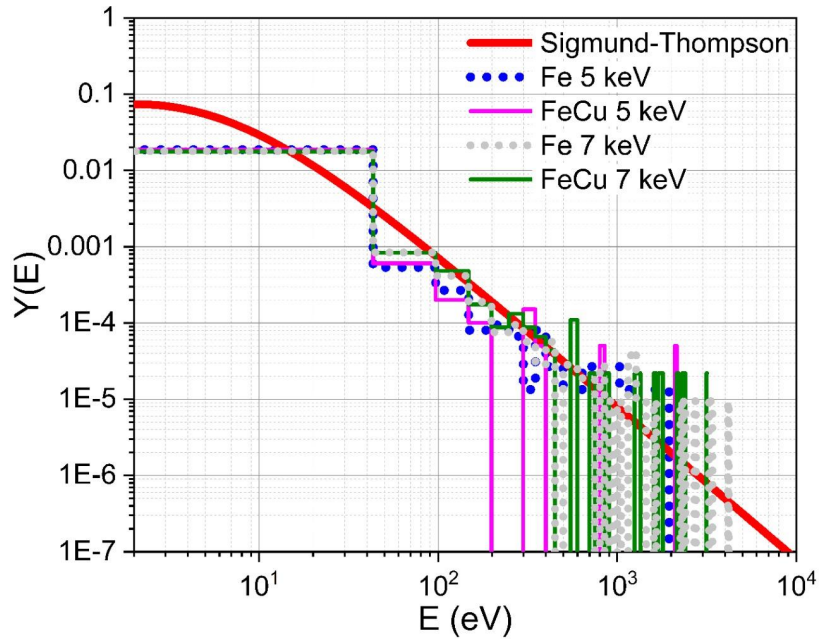

**Fig. S9:** Normalized histogram for the kinetic energy spectra of sputtered atoms as determined by MD simulations for 5 and 7 keV PKA for both Fe and FeCu NP. There is reasonable agreement with the analytical expression from Sigmund-Thompson for binary collision cascades [11], indicating a $E^{-2}$ decay.